\documentstyle[aps,prl,graphics,preprint]{revtex}

\newcommand{\be}{\begin{eqnarray}}
\newcommand{\ee}{\end{eqnarray}}

\begin{document}


\draft \title{Comment on: "Universal Features of Time Dynamics in
a Left-Handed Perfect Lens" by G. Gomez-Santos}
\author{N. Garcia (1)  and M. Nieto-Vesperinas (2)}

\address{(1)E-address: nicolas.garcia@fsp.csic.es, Laboratorio de Fisica de sistemas
Peque\~nos,  Consejo superior de Investigaciones
Cientificas, Serrano 144, Madrid 28006, Spain. \\
(2)E-address: mnieto@icmm.csic.es, Instituto de Ciencia de
Materiales de Madrid, Consejo Superior de Investigaciones
Cientificas, Campus de Cantoblanco, Madrid 28049, Spain}
\maketitle
Recently, we have shown \cite{1} the  impossibility of
perfect lensing with left-handed material slabs, due to the
divergence of the wavefield norm, and hence to the infinite energy
that would accumulate the slab  in order to build the stationary
state. Pendry \cite{2} has subsequently tried to remove this
divergence by calling to small losses in the medium. As proven in
our reply \cite{3}, this does not yield the perfect lens. Then,
Haldane \cite{4} has proposed a cut-off of the wavevector in the
effective medium solution to Maxwell equations by resorting to the
crystal inner  structure of these so-far developed  LHMs. However,
he has overlooked that then, since  a negative refractive index
would be attained in such structure only close to resonant
wavelengths, his definition of group velocity is meaningless
\cite{5}, in addition, as we have shown \cite{6} his cut-off
condition would involve as many as $10^{1000}$ photons in the
slab!.

Now, one more proposal to remove that divergence is reported by
Gomez-Santos \cite{7}. This time neither structural cut-off nor
absorption is accepted. Now the healer should be the time
variation of the evanescent modes (surface polaritons) excited at
the slab interfaces.

It sounds strange that spatial resolution in static media, with
continuous wave sources, which ultimately is conveyed by
stationary waves, be explained by invoking its time dependence. We
thus point out three flaws in this argumentation that, as we shall
show, implies an infinite time to form an image, just as remarked
in \cite{1}

First, it is stated, quoting Pendry \cite{2}, that the evanescent
modes  carry no energy flux and, therefore, energy balance "seems
not to be at risk".
 This is simply wrong since, as shown in \cite{1}, the  current flowing parallel
  to the slab interfaces exponentially increases without limit both with the k-vector
   $\rho_0$ and with the depth into the slab from the first interface. This situation
  can hardly conserve energy.

Secondly, the expression $\Delta \omega_k\approx\ exp(-2\rho_0
a)$, (cf. Eq.(8) of \cite{7}),  is based on a expansion in
\cite{4} (cf. Eq.(1)) which is unjustified (see \cite{5}) at
frequencies
 where  the effective refractive index becomes negative, since then they are very close to
 a resonant frequency of $\epsilon$ and $\mu$.

Third, the discussion on the time dependence of  the wavefunction,
shown in Eqs.(14)-(16) of \cite{7}, is also incorrect because if
$\Delta\omega_k t \approx 1$ and $\Delta\omega_k \approx
\exp(-2\rho_0 a)$, then, obviously $t=1/ \Delta\omega_k $ and thus
the wavefunction norm behaves as $\exp(2\rho_0 a)$ as the time $t$
arbitrarily increases, hence diverging as $\rho_0$ tends to
infinity. This is seen in Fig.1, where we have plotted Eq.(12) of
\cite{7} with $x=\rho_0 a$. As seen, as $t$ increases, the peak of
the wavefunction $E_{trans}$ at which decay given by condition
(15) of \cite{7} starts, progressively shifts to increasing values
of $\rho_0$. In addition, the maximum of $E_{trans}$ exponentially
grows with $\rho_0$ as $t$ increases and its envelope is precisely
the stationary state solution: $exp(2\rho_0 a)$. This is just the
illustration of our remark in \cite{1} underlining that the time
required to fill an stationary state with infinity norm is
infinite. The "perfect lens" image is formed with such state and
infinite resolution in infinite time; meanwhile, the energy stored
in the slab becomes enough to fuse all particles of the Universe!.
In fact, the author of \cite{7} seems, after all, to recognize
this fact in its paragraph below Eq.(17), thus contradicting
himself.

\centerline{FIGURE CAPTIONS}

\begin{description}
\item Fig. 1:Transmitted wavefunction by a slab of left-handed
material of width $a$, according to Eqs.(12) and (13) of Ref.7, as
a function of the parameter $x=\rho_0 a$ at diferent times t;
where $\rho_0$  is the k-vector

\end{description}


\begin{references}

\bibitem{1} N. Garcia and M. Nieto-Vesperinas, Phys. Rev. Lett.
{\bf 88}, 207403(2002).

\bibitem{2} J.B. Pendry, Phys. Rev. Lett. {\bf 85}, 3966 (2000);
J.B. Pendry, Cond-mat/0206561.

\bibitem{3} N. Garcia and M. Nieto-Vesperinas, Cond-mat/0207413.

\bibitem{4} F.D.M. Haldane, Cond-mat/0206420.

\bibitem{5} K.E. Oughstun and H. Xiao, Phys. Rev. Lett, {\bf 78}, 642
(1997).

\bibitem{6} N. Garcia and M. Nieto-Vesperinas, Cond-mat/0207489.

\bibitem{7} G. Gomez-Santos, Cond-mat/0210283.

\end{references}
\end{document}